# Thresholdless deep and vacuum ultraviolet Raman frequency conversion in H$_2$-filled photonic crystal fiber

Manoj K. Mridha[*], Pooria Hosseini, David Novoa, and Philip St.J. Russell

Max Planck Institute for the Science of Light, Staudtstrasse 2, 91058 Erlangen, Germany
*Corresponding author: manoj.mridha@mpl.mpg.de

Coherent ultraviolet (UV) light has many uses, for example in the study of molecular species relevant in biology and chemistry. Very few if any laser materials offer UV transparency along with damage-free operation at high photon energies and laser power. Here we report efficient generation of deep and vacuum UV light using hydrogen-filled hollow-core photonic crystal fiber (HC-PCF). Pumping above the stimulated Raman threshold at 532 nm, coherent molecular vibrations are excited in the gas, permitting highly efficient thresholdless wavelength conversion in the UV. The system is uniquely pressure-tunable, allows spatial structuring of the out-coupled radiation, and shows excellent performance in the vacuum UV. It can also in principle operate at the single-photon level, when all other approaches are extremely inefficient.

**Main Text:**

Molecular species in biology, photochemistry and medicine *(1)* have outer-shell electronic transitions in the vacuum (VUV) and deep ultraviolet (DUV)—from ~100 nm to ~300 nm. Spectroscopy at these wavelengths requires tunable, compact and spectrally narrow UV light sources. Excimer lasers provide direct UV lasing transitions but are fixed-wavelength, inefficient and deliver poor beam quality. Although sum-frequency generation in nonlinear crystals provides a common alternative *(2)*, to be efficient and tunable it requires stringent phase-matching over a broad range of wavelengths (almost impossible to realize in collinear geometries) along with high pump intensities and good spatial overlap between the interacting fields. Moreover, the wavelength-tunability of such systems is in general restricted because very few if any nonlinear crystals provide low dispersion, high transparency and resistance to photo-induced damage in the DUV/VUV. Although these issues have to some degree been addressed by Raman mixing in wide-bore capillaries and cells filled with gas *(3, 4)*, that technique requires intense pump pulses and phase-matching.

Gas-filled hollow-core photonic crystal fiber (HC-PCF), guiding by anti-resonant-reflection, has emerged as a promising alternative that is free from these restrictions. In addition to providing guidance from the VUV to the mid-infrared *(5, 6)*, these fibers offer ultralong light-matter interaction lengths in a hollow channel only few tens of microns wide, together with pressure-tunable dispersion *(6)*. These fibers have reduced the threshold for stimulated Raman scattering (SRS) by orders of magnitude *(7)*, paving the way for multi-octave Raman combs *(8)* reaching into the VUV *(9)*, and broadband frequency conversion in the near infrared *(10)*.

In this paper we report efficient thresholdless frequency conversion of arbitrary DUV signals in hydrogen *(9-11)*, which has the highest Raman gain and frequency shift (~125 THz for the *Q*(1) vibrational transition) of any gas and is transparent down to the VUV. We used a 40-cm-long kagomé-type HC-PCF ("kagomé-PCF") with a core 22 microns in diameter. When gas-filled, the fiber was pumped with ~3.2 ns pulses at 532 nm, resulting via stimulated Raman scattering (SRS) in generation of a noise-seeded Stokes signal at 683 nm (note that rotational SRS can be disregarded (see Supp.)). The beat-note created by these two optical fields drives a coherence wave ($C_w$) of molecular oscillations that, within their coherence lifetime, can be used for thresholdless phase-matched frequency up- or down-conversion of UV pulses. This is possible because of the special frequency dependence of the refractive index of the modes guided in gas-filled kagomé-PCF (see Supp.).

Figure 1(a) shows the ω-β (frequency-wavevector) curves for the kagomé-PCF at 4 bar and 5.3 bar hydrogen (see Supp.). The arrows mark the $C_w$ four-vectors, the vertical projection being the Raman frequency shift and the horizontal projection the $C_w$ wavevector $\beta_{Cw} = \beta_P - \beta_S$, where $\beta_P$ and $\beta_S$ are the propagation constants of the pump and Stokes modes. This $C_w$ can be used for thresholdless up-shifting of a DUV mixing pulse at 266 nm (propagation constant $\beta_P^m$) to its anti-Stokes band (239 nm, propagation constant $\beta_{AS}^m$) and to greatly lower the threshold for down-shifting to the Stokes band (299 nm, propagation constant $\beta_S^m$), provided the dephasing rates $\Delta\beta_{AS} = |\beta_{AS}^m - \beta_P^m - \beta_{Cw}|$ and $\Delta\beta_S = |\beta_P^m - \beta_S^m - \beta_{Cw}|$ are such that the dephasing lengths $\pi/\Delta\beta_{AS}$ and $\pi/\Delta\beta_S$ are longer than the fiber length. Fig. 1(a) shows that these dephasing rates can be made vanishingly small (i.e., $\Delta\beta_i \to 0$) for collinear generation of anti-Stokes and Stokes signals at ~4 and ~5.3 bar, illustrating the exquisite pressure-tunability of these conversion processes. Scanning electron microscopy of the fiber cross-section (Fig. 1(b)) revealed that the core-wall thickness was ~96 nm, resulting in a loss-inducing anti-crossing *(12)* at 225±5 nm. Away from this loss band the fiber has low loss and spectrally flat dispersion. The 22 µm core diameter ensures that the optimum dispersion landscape for UV dynamics is achieved at pressures well above 1 bar, something that is impossible to achieve with wide-bore capillaries or in bulk gas cells *(3, 4)*.

The experimental set-up is sketched in Fig. 1(c). The linearly polarized pump pulses and DUV mixing pulses (duration ~3 ns) were co-launched into the fundamental mode of the fiber (Supp.). The mixing pulse energy was kept well below the SRS threshold, ensuring that the dynamics were driven purely by the pre-existing molecular coherence. Since we operated in the "transient" regime *(13)*, the molecular coherence builds up under the pump pulse envelope. As a result, the efficiency could be optimized by tuning the mixing pulse delay; a value of ~1 ns was optimal. The gas pressure was regulated in fine steps (~50 mbar) and the generated UV bands spatially separated using a prism. Figure 1(d) shows the spectrum of the dual comb when ~27 µJ green pulse energy and ~70 nJ DUV mixing energy were launched into the fiber at ~6 bar pressure. With the 532 nm light blocked, no Raman sidebands of the mixing signal were observed, as expected (see Fig. 1(d)). The mixing beam energy was chosen so that all the bands could be detected with the equipment available. The technique should, however, work down to the single-photon limit.

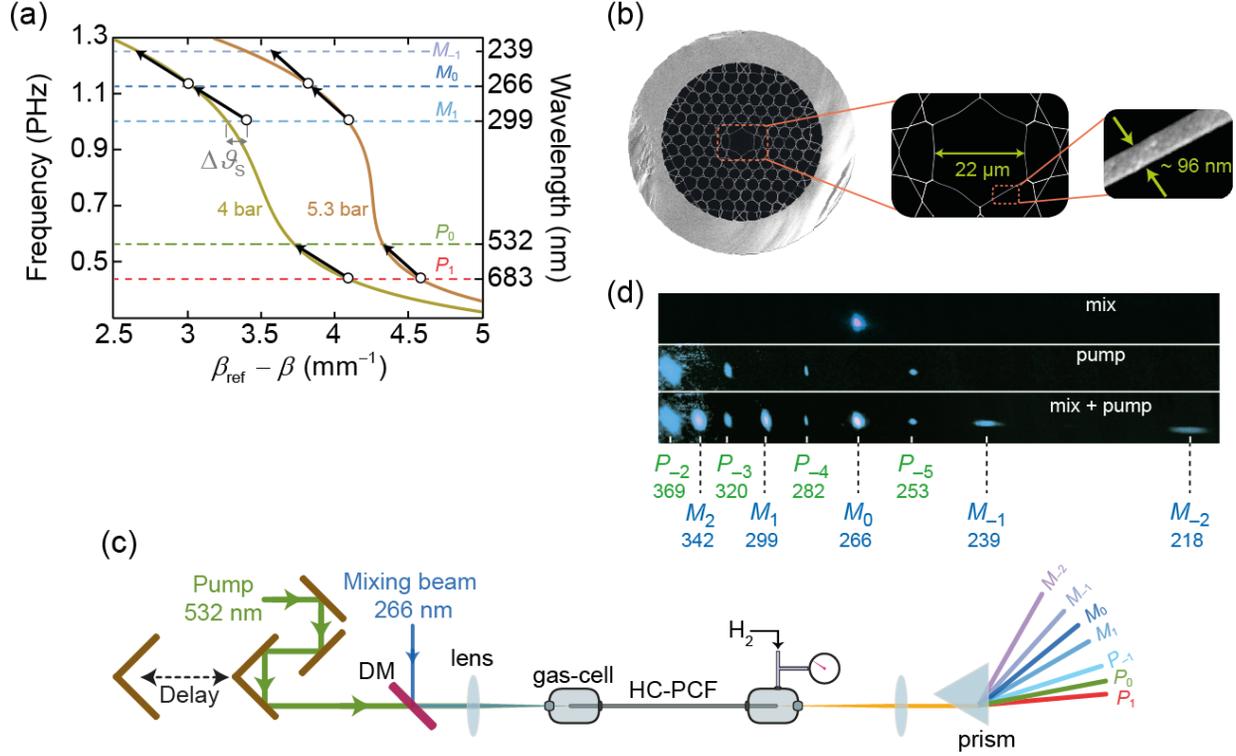

Figure 1. (a) Dispersion curves (neglecting loss bands caused by anti-crossings between the core mode and modes in the glass walls surrounding the core) for the $LP_{01}$ mode of the HC-PCF filled with hydrogen at two different pressures. The subtle features of the curves are magnified by plotting frequency against ($\beta_{ref} - \beta$), where $\beta_{ref}$ is a linear function of frequency, chosen such that ($\beta_{ref} - \beta$) is zero for the $LP_{01}$ mode at 1650 THz. The arrows represent the coherence waves excited by the beating of the green pump and its Stokes band at 683 nm. (b) Scanning electron micrographs of the kagomé-PCF microstructure. (c) Schematic of the experimental set-up. DM: dichroic mirror. (d) Dispersed spectra cast on a screen for three different cases: mixing signal only (upper); green pump only (middle) along with its Raman bands; and co-launched mixing signal and green pump (lower). The spectral sidebands originating from the green pump and the mixing signal, both indicated along with their wavelengths in nanometers, are easily distinguishable.

In the following we use $M_m$ to denote the energy in the $m$-th sideband of the mixing signal at the fiber output ($m > 0$ for Stokes bands) and $P_m$ the energy in the $m$-th sideband of the green pump at the fiber output. $P_{00}$ is the launched green energy.

Figure 2(a) shows the pressure dependence of the overall conversion efficiency to the $M_m$ sidebands, quantified by $\eta_M = 1 - M_0/M_{00}$ where $M_{00}$ is the value of $M_0$ with the green pump light switched off. The launched pump energy was ~10 µJ and the energy of the 266 nm pulse alone, measured at the fiber output, was ~100 nJ. Several local maxima in $\eta_M$ are apparent in Fig. 2(a), reaching peak values greater than 45%. We attribute the complex pressure dependence of $\eta_M$, which is particularly noticeable at higher pump energies (Supp.), to several factors. Firstly, phase-matching to $M_{-1}$ and $M_1$ is satisfied at different pressures (see Fig. 1(a)). Secondly, the scattering process is enhanced at higher pressures through increased Raman gain (*14*). Finally, as the Raman gain increases the first-order sidebands become stronger, resulting in conversion to higher-order sidebands. Higher values of $P_{00}$ also increased $\eta_M$; in the experiment, raising $P_{00}$ to ~27 µJ resulted in $\eta_M \sim 60\%$ (see Supp.). All these results are corroborated by numerical solutions of multimode Maxwell-Bloch equations (see Fig. 2(a) lower and Supp. for details).

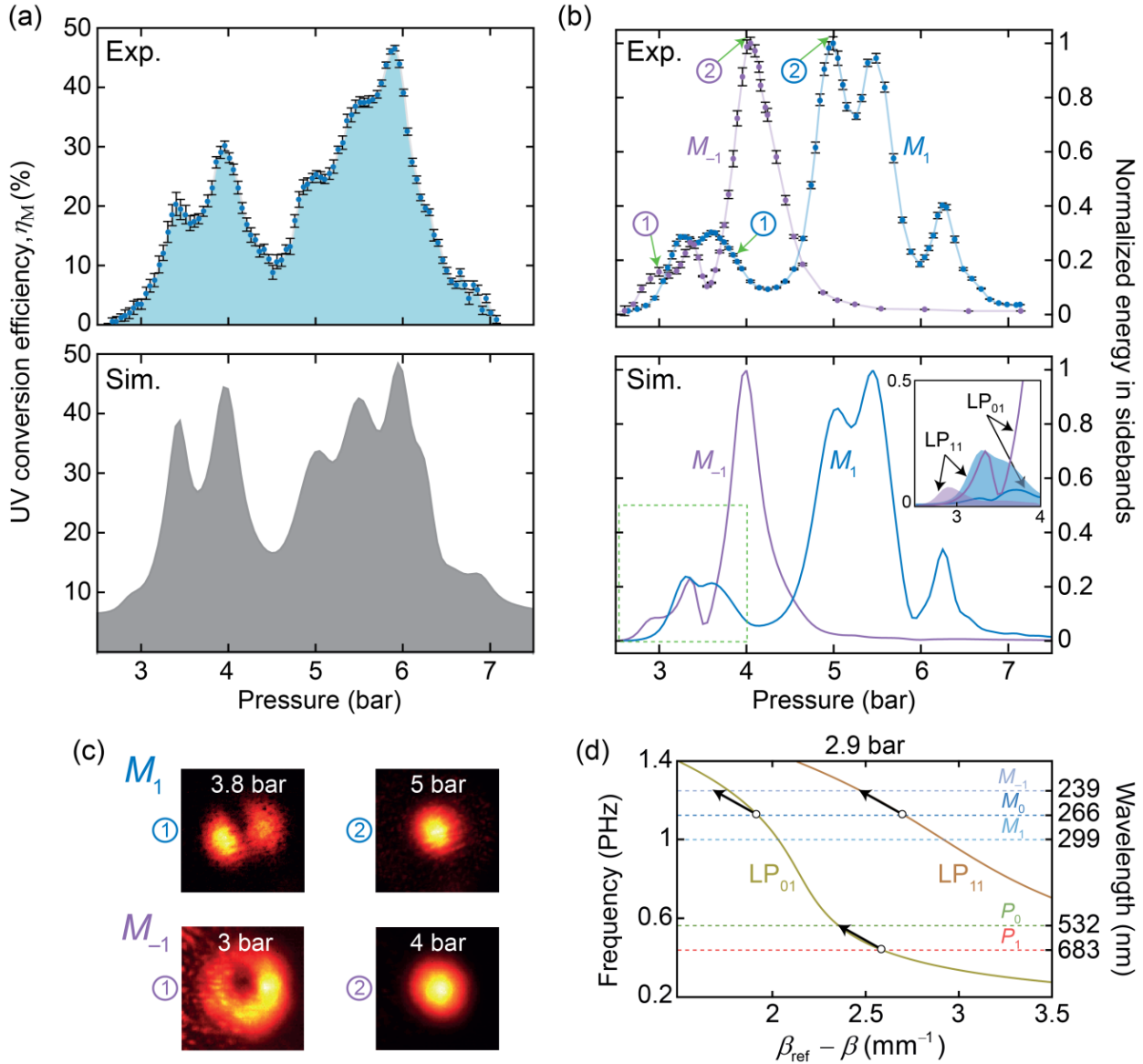

Figure 2. Experimental (upper) and numerically simulated (lower) pressure dependence of (a) overall conversion efficiency to sidebands, $\eta_M$, (b) $M_{-1}$ (239 nm) and $M_1$ (299 nm). The inset in the lower part of (b) shows the fundamental and HOM content of the $M_{-1}$ and $M_1$ signals at low pressure. The simulations suggest that the launched mixing signal $M_{00}$ contained ~30% of the $LP_{11}$ mode. (c) Far-field transverse intensity profiles of the $M_1$ and $M_{-1}$ beams, showing their emergence in the $LP_{01}$-like (right) and $LP_{11}$-like mode (left). (d) Dispersion curves for the $LP_{01}$ and $LP_{11}$ modes when filled with 2.9 bar of $H_2$. At this pressure, phase-matching occurs between the $M_0$ and $M_{-1}$ signals both being in the $LP_{11}$ mode.

The pressure dependence of the $M_{-1}$ and $M_1$ signals, normalized to their peak values, is shown in Figure 2(b). There is remarkably good agreement between theory and experiment. Fig. 1(a) shows that both $M_{-1}$ and $M_1$ peak close to their predicted phase-matching pressures (the double-humped structure arises from the dynamics of the conversion process). When $M_{-1}$ is highest, $M_1$ is very weak and vice-versa, demonstrating full selectivity of the direction of energy exchange between

$M_0$ and the coherence wave. The dips in signal at ~3.5 bar for $M_{-1}$ and ~6 bar for $M_1$ are caused by conversion to the next-order sidebands $M_{-2}$ (218 nm) and $M_2$ (342 nm) (see Supp. for details). The revival of the signals at low gas pressure is caused by the presence of higher-order modes (HOMs), as seen in the simulations (inset in Fig. 2(b)) and confirmed by the far-field images in Fig. 2(c). These frequency-shifted HOMs result from some HOM content in the mixing signal $M_{00}$, together with efficient phase-matched transitions to ultraviolet HOMs via intramodal coherence waves (Fig. 2(d)). Since the UV wavelength conversion is mode-selective, it can be tailored to generate specific DUV-VUV beam profiles for different applications.

To demonstrate that the conversion process is thresholdless, we recorded $\eta_M$ as a function of $M_{00}$ at 4 bar (see Fig. 3(a)), revealing that $\eta_M$ is almost independent of $M_{00}$, and that the scattering is effectively linear. This means that the system should in theory be efficient down to the single-photon limit. In practice, the lowest values of $M_{00}$ that could be measured in the experiments were ~5 nJ, limited by the detector sensitivity. At high DUV energies, however, the process becomes nonlinear because the mixing beam starts to generate its own molecular coherence via SRS.

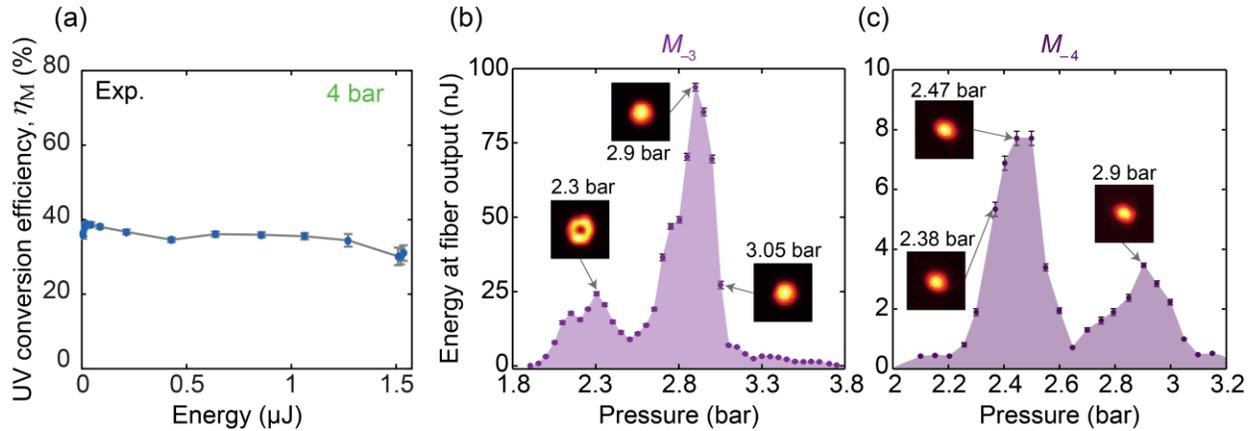

Figure 3 (a) Overall sideband conversion efficiency $\eta_M$ plotted against $M_{00}$ for a green pump energy $P_{00}$ = 12.6 µJ. (b) Pressure dependence of the $M_{-3}$ signal (199 nm). (c) Pressure dependence of the $M_{-4}$ signal (184 nm). See the text for experimental parameters. The far-field spatial profiles of the mixing beam sidebands are shown in the insets.

As suggested above, the unique guiding properties of gas-filled kagomé-PCF means that frequency conversion will also work in the VUV. Figs. 3(b) and 3(c) show the pressure-dependence of the $M_{-3}$ (199 nm) and $M_{-4}$ (184 nm) signals for $P_{00}$ ~ 29 µJ and $M_{00}$ ~ 1.43 µJ. At ~2.9 bar pressure the $M_{-3}$ signal reaches ~94 nJ (see Fig. 3(b)), corresponding to a conversion efficiency of 6.6 % from the $M_{00}$ signal. It is likely that the efficiency can be further increased by optimizing the system—far from a trivial task, given the onset of the $M_{-4}$ signal (lower peak in Fig. 3(c)) and the complex spatio-temporal evolution of the coherence at high pump energies (Supp.). For example, we found that by reducing $P_{00}$ to ~20 µJ the conversion efficiency to the $M_{-3}$ signal rose to ~9.2 % (see Supp.). At ~2.4 bar, ~8 nJ of VUV light was generated in the $M_{-4}$ sideband at 184 nm, corresponding to ~0.6 % conversion efficiency from $M_{00}$ signal, see Fig. 3(c). It is remarkable that, apart from the low-pressure region of Fig. 3(b), the modal selectivity of the system makes it possible to generate VUV radiation in a pure $LP_{01}$-like mode (see insets of Fig. 3(b) and 3(c)), something that was not possible in previous work on noise-seeded Raman combs (9).

In conclusion, long-lived molecular coherence excited in the gas-filled core of a HC-PCF enables highly-efficient, pressure-tunable frequency conversion of arbitrary signals in the DUV and VUV.

The modal content of the DUV-VUV bands can be controlled to a great degree, making it possible to generate both Gaussian-like and spatially-structured beams. We anticipate that, with further improvements in the DUV-VUV performance of broadband HC-PCFs, a family of unique high-performance coherent UV light sources will emerge, with applications in, for example, generation of arbitrary waveforms (*15*), tailored attosecond pulse trains (*16*) and UV frequency combs (*17*), without need for high energy pump pulses.

**References:**

(1) M. T. N.-Petersen, G. P. Gajula, S. B. Petersen, "UV Light Effects on Proteins: From Photochemistry to Nanomedicine" in *Satyen Saha, editors. Molecular Photochemistry*, (IntechOpen). Doi 10.5772/37947.

(2) H. Xuan, H. Igarashi, S. Ito, C. Qu, Z. Zhao, Y. Kobayashi, High-power, solid-state, deep ultraviolet laser generation. *Appl. Sci.* **8**, 233 (2018).

(3) Y. Mori, T. Imasaka, Generation of ultrashort optical pulses in the deep-ultraviolet region based on four-wave Raman mixing. *Appl. Sci.* **8**, 784 (2018).

(4) D. Vu, T. N. Nguyen, T. Imasaka, Generation of a femtosecond vacuum ultraviolet optical pulse by four-wave Raman mixing. *Opt. Laser Technol.* **88**, 184-187 (2017).

(5) C. Wei, R. J. Weiblen, C. R. Menyuk, J. Hu, Negative curvature fibers. *Adv. Opt. Photon.* **9**, 504–561 (2017).

(6) P. St.J. Russell, P. Hölzer, W. Chang, A. Abdolvand, J. C. Travers, Hollow-core photonic crystal fibres for gas-based nonlinear optics. *Nat. Photonics* **8**, 278–286 (2014).

(7) F. Benabid, J. C. Knight, G. Antonopoulos, P. St.J. Russell, Stimulated Raman scattering in hydrogen-filled hollow-core photonic crystal fiber. *Science* **298**, 399–402 (2002).

(8) F. Couny, F. Benabid, P. J. Roberts, P. S. Light, M. G. Raymer, Generation and photonic guidance of multi-octave optical-frequency combs. *Science* **318**, 1118–1121 (2007).

(9) M. K. Mridha, D. Novoa, S. T. Bauerschmidt, A. Abdolvand, P. St.J. Russell, Generation of a vacuum ultraviolet to visible Raman frequency comb in $H_2$-filled kagomé photonic crystal fiber. *Opt. Lett.* **41**, 2811–2814 (2016).

(10) S. T. Bauerschmidt, D. Novoa, A. Abdolvand, P. St.J. Russell, Broadband-tunable $LP_{01}$ mode frequency shifting by Raman coherence waves in a $H_2$-filled hollow-core photonic crystal fiber. *Optica* **2**, 536–539 (2015).

(11) K. Hakuta, M. Suzuki, M. Katsuragawa, J. Z. Li, Self-induced phase matching in parametric anti-Stokes stimulated Raman scattering. *Phys. Rev. Lett.* **79**, 209–212 (1997).

(12) J. L. Archambault, R. J. Black, S. Lacroix, J. Bures, Loss calculations for antiresonant waveguides. *J. Lightwave Technol.* **11**, 416−423 (1993).

(13) P. Hosseini, D. Novoa, A. Abdolvand, P. St.J. Russell, Enhanced control of transient Raman scattering using buffered hydrogen in hollow-core photonic crystal fibers. *Phys. Rev. Lett.* **119**, 253903 (2017).

(14) W. K. Bischel and M. J. Dyer, Wavelength dependence of the absolute Raman gain coefficient for the *Q*(1) transition in $H_2$. *J. Opt. Soc. Am. B* **3**, 677 (1986).

(15) H.-S. Chan, Z. -M. Hsieh, W. -H. Liang, A. H. Kung, C. -K. Lee, C. -J. Lai, R. -P. Pan, L. -H. Peng, Synthesis and measurement of ultrafast waveforms from five discrete optical harmonics. *Science* **331**, 1165-1168 (2011).

(16) K. Yoshii, J. K. Anthony, M. Katsuragawa, The simplest route to generating a train of attosecond pulses. *Light Sci. Appl.* **2**, e58 (2013).

(17) G. Porat, C. M. Heyl, S. B. Schoun, C. Benko, N. Dörre, K. L. Corwin, J. Ye, Phase-matched extreme-ultraviolet frequency-comb generation. *Nat. Photonics* **12**, 387-391 (2018).

(18) M. A. Finger, N. Y. Joly, T. Weiss, P. St. J. Russell, Accuracy of the capillary approximation for gas-filled kagomé-style photonic crystal fibers. *Opt. Lett.* **39**(4), 821–824 (2014).

(19) S. T. Bauerschmidt, D. Novoa, P. St.J. Russell, Dramatic Raman gain suppression in the vicinity of the zero dispersion point in a gas-filled hollow-core photonic crystal fiber. *Phys. Rev. Lett.* **115**, 243901 (2015).

(20) G. P. Agrawal, *Nonlinear Fiber Optics* (Academic Press, San Diego, ed. 4, 2007).

(21) E. R. Peck, S. Huang, Refractivity and dispersion of hydrogen in the visible and near infrared. *J. Opt. Soc. Am.* **67**, 1550 (1977).

(22) B. M. Trabold, D. Novoa, A. Abdolvand, P. St.J. Russell, Selective excitation of higher order modes in hollow-core PCF via prism-coupling. *Opt. Lett.* **39**(13), 3736–3739 (2014).

Supplementary Materials for

**Thresholdless deep and vacuum ultraviolet Raman frequency conversion in $H_2$-filled photonic crystal fiber**

**This PDF file includes:**

    Materials and Methods
    Supplementary Text
    Figs. S1 to S7
    Table S1
    References 1 to 6

**Materials and Methods**

Laser source

We employed an injection-seeded Nd:YAG laser delivering 1064 nm, 3.2 ns pulses at 3 kHz repetition rate. By frequency-doubling it in a Potassium-Titanyl-Phosphate (KTP) crystal, we obtained the green pump source at 532 nm. The 266 nm mixing beam was then obtained by another frequency-doubling stage of the 532 nm pump in a Beta-Barium-Borate (BBO) crystal.
An uncoated plano-convex $CaF_2$ lens with focal length 100 mm was used to launch both the pump and the mixing beam into the fiber. To mitigate the in-coupling mismatch of the two wavelengths due to chromatic aberration, we used two telescopes placed in the optical paths of both beams—before the dichroic mirror (see Fig. (1c) in main text). The use of this arrangement resulted in independent control of the beam diameters of both beams, which facilitated their simultaneous coupling into the fiber.

Gas system

The HC-PCF is placed in two 8-cm-long gas-cells connected by a hollow metallic tube. Light enters and exists the gas-cells through 3 mm thick and 10 mm in diameter $MgF_2$ optical windows. The pressure in the gas-cells was manually regulated with a very fine step size of ~ 50 mbar. This was made possible by the combined action of a number of gas-flow components: a coarse self-venting pressure regulator, a needle valve and a metering valve. The metering valve, when completely closed, played a key role in providing the minimum gas flow with additional control being provided by the needle valve.

**Supplementary Text**

Dispersion landscape of kagomé-type HC-PCF

The wavelength-dependent propagation constant $\beta_{ij}(\lambda)$ for the LP$_{ij}$-like modes of a gas-filled kagomé-style hollow-core photonic crystal fiber is analytically calculated using the modified Marcatili-Schmelzer model *(18)*

$$\beta_{ij}(\lambda) = k_0 \sqrt{n_{gas}^2(p,\lambda) - \lambda^2 u_{ij}^2 / (2\pi a)^2} \quad \text{(S1)}$$

Where $u_{ij}$ is the $j^{th}$ root of the $i^{th}$ order Bessel function of the first kind, with ($i$, $j$) being also the azimuthal and radial mode orders, $k_0 = 2\pi / \lambda$ is the vacuum wavevector, $n_{gas}(p,\lambda)$ is the pressure, $p$, and wavelength-dependent refractive index of the filling gas, and $a$ is the area-preserving core radius. Note that this model disregards the influence of loss bands caused by anti-crossings between the core mode and modes in the glass walls surrounding the core, which are irrelevant in our experimental conditions as we have discussed in the main text.

## Set of coupled Maxwell-Bloch equations involving multiple fiber modes

The evolution of the electric fields of the Raman sidebands, as well as the molecular coherence triggered by the green pump are modeled through a set of coupled Maxwell-Bloch equations involving several fiber modes *(19)*

$$\frac{\partial}{\partial z}E_{\sigma,l} = -\kappa_{2,l}\frac{\omega_l}{\omega_{l-1}}\sum_{\nu\xi\eta}^{M}i\frac{s_{\sigma\nu\xi\eta}}{S_{\nu\xi}}Q_{\nu\xi}E_{\eta,l-1}r_{\eta,l-1}r^*_{\sigma,l}$$

$$-\kappa_{2,l+1}\sum_{\nu\xi\eta}^{M}i\frac{s_{\sigma\nu\xi\eta}}{S_{\nu\xi}}Q^*_{\nu\xi}E_{\eta,l+1}r_{\eta,l+1}r^*_{\sigma,l}$$

$$-\frac{1}{2}\alpha_{\sigma,l}.$$
(S2)

$$\frac{\partial}{\partial\tau}Q_{\nu\xi} = -Q_{\nu\xi}/T_2 - i(S_{\nu\xi}/4)\sum_{l}\kappa_{1,l}E_{\nu,l}E^*_{\xi,l-1}r_{\nu,l}r^*_{\xi,l-1},$$ (S3)

where $l$ is an integer and denotes the Raman sidebands of the pump with frequencies $\omega_l = \omega_P + 2\pi l\Omega_R$, where $\omega_P$ is the pump frequency and $\Omega_R$ is the Raman frequency shift. The summation in the equation goes over the permutation of all the modal sets $M$. The complex electric field amplitude of a given mode $\sigma$ is defined as $e_{\sigma,l}(z,\tau) = F_{\sigma,l}(x,y)E_{\sigma,l}(z,\tau)r_{\sigma,l}$, where $E_{\sigma,l}(z,\tau)$ is the slowly-varying field envelope, $F_{\sigma,l}(x,y)$ is the normalized transverse spatial profile and $r_{\sigma,l} = \exp[-i\beta_\sigma(\omega_l)]$. $\alpha_{\sigma,l}$ represent the propagation losses for the fields and $Q_{\nu\xi}$ denotes the amplitude of both intermodal and intramodal coherence waves. The equations are derived assuming that most of the molecules remain in their ground state. In addition we consider that all quasi-monochromatic fields propagate at the speed of light $c$, and we define a co-moving time frame such that $\tau = t - z/c$, where $t$ is the absolute time. $\kappa_{1,l}$ and $\kappa_{2,l}$ are coupling constants given by:

$$\kappa_{1,l} = \sqrt{\frac{2\gamma_l c^2 \varepsilon_0^2}{NT_2\hbar\omega_{l-1}}}, \qquad \kappa_{2,l} = \frac{N\hbar\omega_{l-1}\kappa_{1,l}}{2\varepsilon_0 c}$$ (S4)

where $T_2$ is the dephasing time of the Raman polarization and is linked to the linewidth $\Delta\nu$ of the Raman transition through the relation $T_2 = 1/\pi\Delta\nu$, $\hbar$ is the reduced Planck's constant, $\varepsilon_0$ is the vacuum permittivity and $N$ is the molecular number density. The values of these coefficients are obtained from experimental measurements of the material Raman gain $\gamma_l$. The generalized nonlinear spatial overlap integrals $S_{\nu\xi}$ and $s_{\sigma\nu\xi\eta}$ are given by *(20)*:

$$s_{\sigma\nu\xi\eta} = \frac{\int F_\sigma^* F_\nu^* F_\xi F_\eta dA}{\left(\int |F_\sigma|^2 dA \int |F_\nu|^2 dA \int |F_\xi|^2 dA \int |F_\eta|^2 dA\right)^{1/2}} A_{eff} \qquad s_{\nu\xi} = \frac{\int |F_\nu|^2 |F_\xi|^2 dA}{\left(\int |F_\nu||F_\xi| dA\right)^2} A_{eff} \qquad (S5)$$

where $A_{eff}$ is the effective mode area of the pump in the fundamental mode and the integral is calculated over the transverse cross-section of the fiber. $F_{\sigma,l}(x, y)$ are obtained from finite element calculations in an ideal kagomé structure and assumed to be wavelength-independent for all the Raman bands relevant to this work.

Once the molecular coherence is calculated, the evolution and scattering of a weak mixing field $E_{\sigma, mix}$ and frequency $\omega_{mix}$ to sidebands denoted by an integer $m$ such that $\omega_m = \omega_{mix} + 2\pi m \Omega_R$ can be described by the following equation:

$$\frac{\partial}{\partial z} E_{\sigma, m} = -\kappa_{2,m} \frac{\omega_m}{\omega_{m-1}} \sum_{\nu\xi\eta}^{M} i \frac{s_{\sigma\nu\xi\eta}}{S_{\nu\xi}} Q_{\nu\xi} E_{\eta, m-1} r_{\eta, m-1} r_{\sigma, m}^*$$

$$- \kappa_{2,m+1} \sum_{\nu\xi\eta}^{M} i \frac{s_{\sigma\nu\xi\eta}}{S_{\nu\xi}} Q_{\nu\xi}^* E_{\eta, m+1} r_{\eta, m+1} r_{\sigma, m}^*$$

$$- \frac{1}{2} \alpha_{\sigma, m} \qquad (S6)$$

Note that this model is valid in the limit of weak mixing signals, i.e. when they are not strong enough to excite molecular coherence waves on their own, thereby inducing any back action on the pre-existing coherence.

Parameters used in the simulations

As in the experiment, we considered a fiber length of 40 cm and core radius $a = 22$ µm. The material dispersion of hydrogen gas is adopted from *(21)*. Based on the experimental observations, only the LP$_{01}$-like and LP$_{11}$-like modes were considered for the pump, mixing beam and their sidebands. To calculate the spatio-temporal evolution of the coherence waves in the system, we included three vibrational Stokes and anti-Stokes bands. For the green pump and its first Stokes band at 683 nm, we included measured loss values of 0.96 dB/m and 1.94 dB/m, respectively. Due to the lack of loss measurements for the remaining Raman lines of the green pump and the mixing beam, their loss coefficients were free parameters, reasonably adjusted to get close agreement to the experiments (see Table S1), although small variations of these values do not produce any significant changes in the observed dynamics.

For the LP$_{11}$-like mode, the losses of the different lines were simply double of those of the corresponding fundamental mode, an approximation that has been experimentally validated for similar fiber structures *(22)*.

The mixing pulse was delayed by 1 ns with respect to the pump. The green pump was launched only in the fundamental mode. To obtain the numerical results shown in the main paper, we considered the initial injection of 100 nJ and 40 nJ of the mixing beam respectively in the $LP_{01}$ and $LP_{11}$-like mode. To model the initial stages of Raman amplification and subsequent ultraviolet scattering, a noise floor of 50 V/m was considered for all the Raman sidebands in both the modes, as well as for the pump in the $LP_{11}$-like mode. Owing to the uncertainty in the measured Raman gain values $\gamma_l$ at low pressures *(14)*, the best agreement with the experiments involving 10 µJ of green pump energy launched in the fiber was obtained by considering 7 µJ of pump energy in the computations. As discussed above, for mixing pulse energies exceeding 1 µJ, our simplified numerical model for the dual-pump scheme is no longer valid as the back-action of the mixing beam on the generated molecular coherence cannot be disregarded.

## Coherence wave generated mainly from pump (532 nm) to Stokes (683 nm) transition

The beat-note created by two strong adjacent Raman sidebands is responsible for the excitation of coherence waves. In all experimental spectra displayed in Fig. S1, the first vibrational Stokes (*$P_1$*) is the strongest among all the vibrational Raman lines—only ~ 4 dB weaker than the pump. In addition, the rotational lines (separated by ~18 THz) are much weaker than the vibrational sidebands due to their lower gain, meaning that the generated coherence waves are mainly driven by the $P_0 - P_1$ Raman transition. It is important to note that not only rotational SRS lines can influence the molecular coherence, but also higher-order vibrational Stokes sidebands can play a role. At low pressures, $P_2$ is more than 15 dB lower than $P_1$ even at high pump energies (see Fig. S1 (a)-(b)). Figures S1 (c-h) show the experimental output spectra recorded for two different pump energies at three different pressures of hydrogen. The number of Raman lines and their strength grows with increasing pressures and large pump energies—something expected as the Raman gain for hydrogen increases with pressure. Irrespective of the pressures shown, at relatively low pump energy of ~11 µJ, the $P_2$ signal is more than 30 dB weaker than the $P_1$. However, for pump pulse energies of ~25 µJ, the $P_2$ signal grows significantly, and is less than 5 dB lower than the $P_1$ signal. In these conditions, the coherence wave generated from $P_1$-to- $P_2$ transitions, although weaker, may influence the efficiency of the mixing beam depletion.

## Influence of the pump energy on the ultraviolet conversion efficiency

The complexity in the pressure-dependent conversion efficiency of the mixing beam increases with increasing pump pulse energy. This is verified experimentally in Fig. S2 and S3 and numerically in Fig. S4.
   Figure S2 plots the mixing beam conversion efficiency at four different pressures as a function of the green pump energy. For all the pressures, the efficiency increases with pump energy due to the presence of stronger coherence waves. However, probably due to the involved phase relation between the coherence waves and the interacting ultraviolet signals, there is a slight drop in the efficiency at moderate pump energies, followed by a revival. Such behavior has also been corroborated by numerical simulations (not shown).
   Figure S3 displays the UV conversion efficiency for a launched green pump energy of ~18 µJ, larger than that employed in the experiments described in Fig. 2(a) in the main text. Comparing these two figures, there is a clear increase in complexity of the dynamics of the system as the gas pressure is varied.

Figure S4 plots the numerical results of the in-fiber dynamics of the mixing beam and the growth of its sidebands ($M_{-2}$ to $M_2$) for two different pump energies at two different pressures. In Fig. S4(a, b), the $M_{-1}$ signal is the strongest among all the sidebands; the same occurs for $M_2$ in Fig. S4(c). This constitutes a further numerical verification of the phase-matched generation process discussed in the main text and is also supported by Fig. S5. Irrespective of the pressure, it can be seen from these figures that on increasing the pump energy, the generation of the sidebands reaches almost saturation in a relatively short fiber length. In addition, beyond this short distance within the fiber, all these lines oscillate and dynamically exchange energy over the remaining fiber length. These interactions make it difficult to accurately predict a priori the complex profile of the UV conversion efficiency recorded at the fiber output—as can be experimentally observed in Fig. S3.

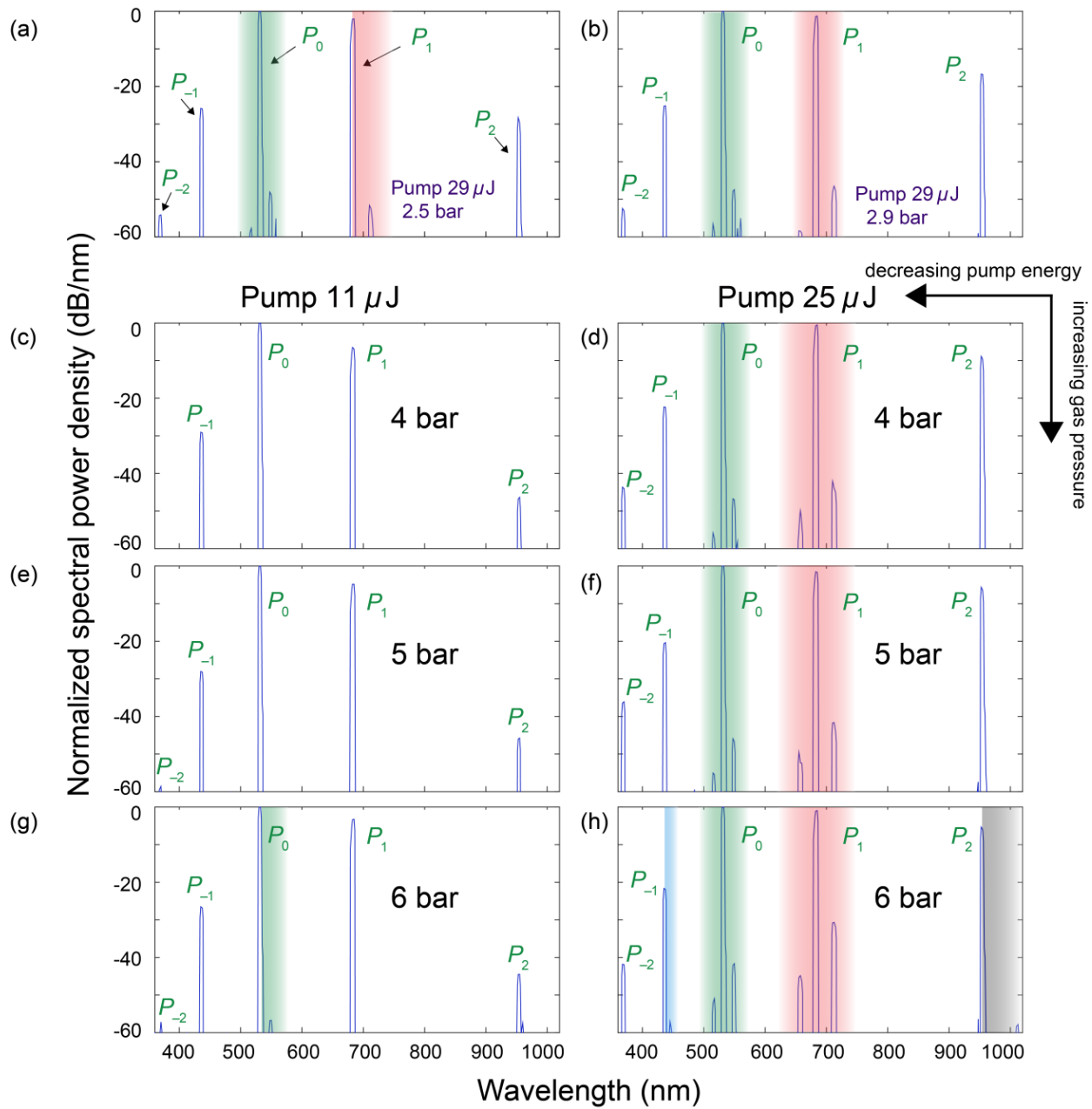

**Fig. S1.**
**Spectra recorded at the output of the gas-filled HC-PCF for different gas pressures and launched pump energies**. The spectra consist of a ro-vibrational Raman comb of the pump and the shaded regions enclose the rotational lines of the respective vibrational line or the pump. For each spectrum, all sidebands are normalized to the peak pump power.

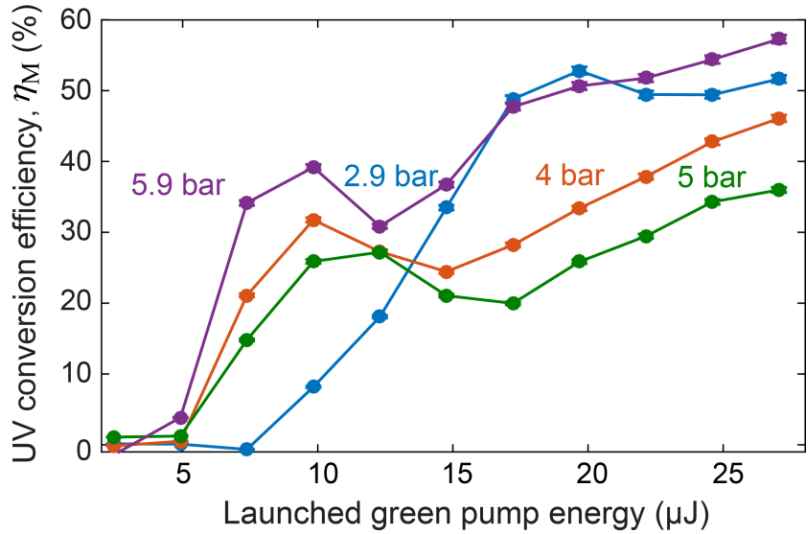

**Fig. S2.**

**Experimental observation of the energy dependence of the ultraviolet conversion efficiency.** We recorded the conversion efficiency the 266 nm mixing beam at four different pressures for increasing green pump energies. Although the maximum attainable efficiency was not reached (i.e. the curves are not yet saturated), values around ~60% can readily be achieved at the highest energies launched in the fiber.

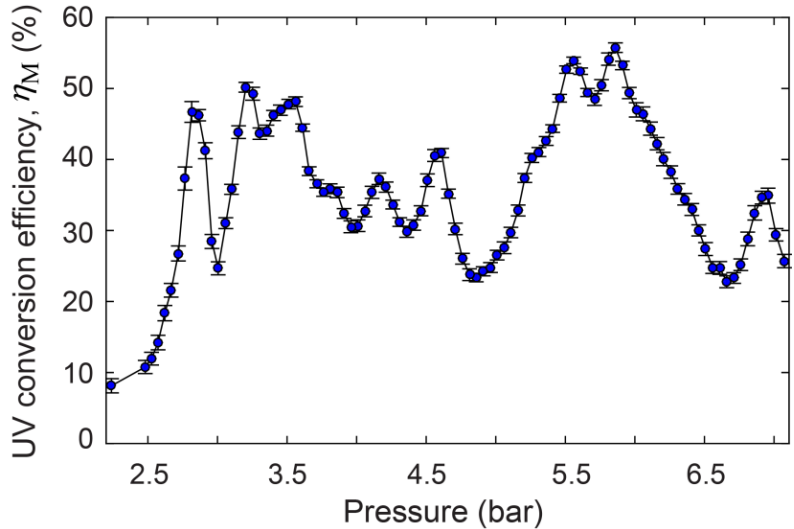

**Fig. S3.**

**Measured UV conversion efficiency for a launched green pump energy of ~18 μJ**. The dynamics of the frequency conversion process becomes more complex for high pump energies (this is to be compared with the results shown in Fig. 2(a) of the main text).

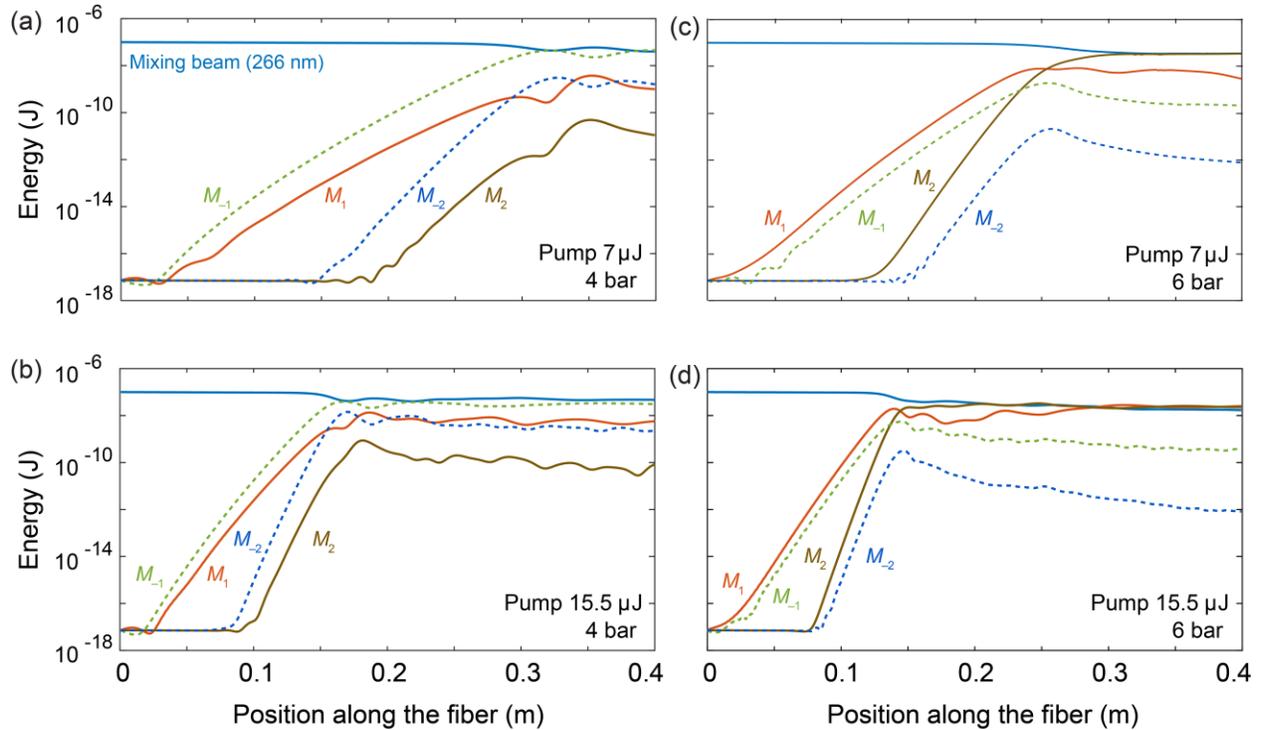

**Fig. S4.**

**Simulated evolution of the LP$_{01}$ mode content of the mixing beam and its higher-order Raman sidebands, ranging from $M_{-2}$ to $M_2$, along the fiber for two different gas pressures and pump energies.** The launched energy of the mixing beam is 100 nJ and, for simplicity, only its LP$_{01}$ mode content is displayed. Individual lines follow the same color code throughout the four sub-figures.

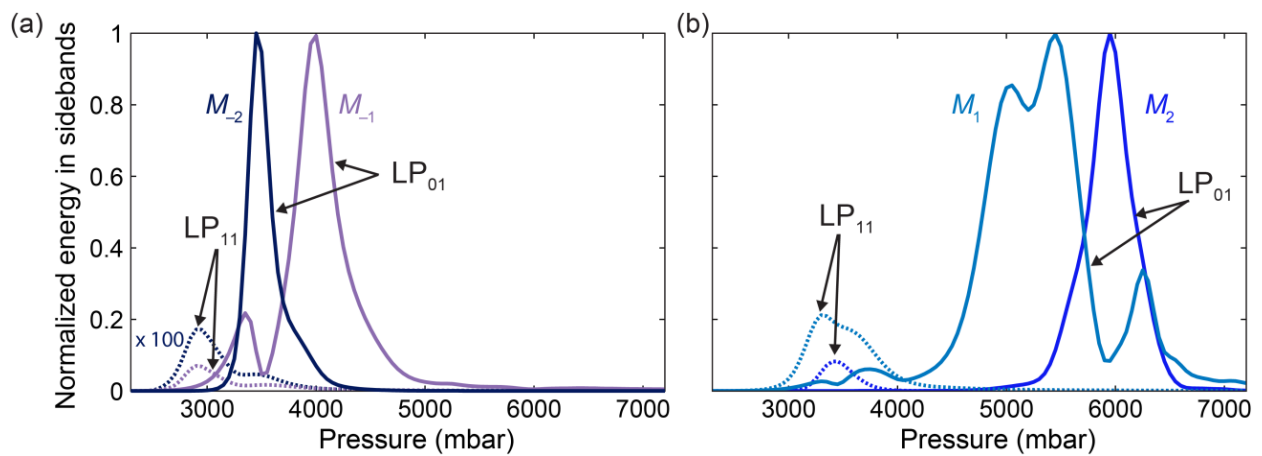

**Fig. S5.**

**Simulated pressure-dependent generation of (a) $M_{-1}$ (239 nm), $M_{-2}$ (218 nm), (b) $M_1$ (299 nm) and $M_2$ (342 nm) for both LP$_{01}$ and LP$_{11}$ mode contributions**. The parameters used in the simulations are the same as in Fig. 2(a)-(b) in the main text. The energy of each individual Raman band is normalized to the peak of its respective $LP_{01}$-like mode content. The $LP_{11}$ mode content of the $M_{-2}$ is magnified by a factor of 100 for clarity. The pressure-dependent strengths of $M_{-2}$ and $M_2$ reach their peaks around 3.5 bar and 6 bar, respectively, which correspond to the valleys displayed in the measurements of the $M_{-1}$ and $M_1$ signals, as mentioned in the main text.

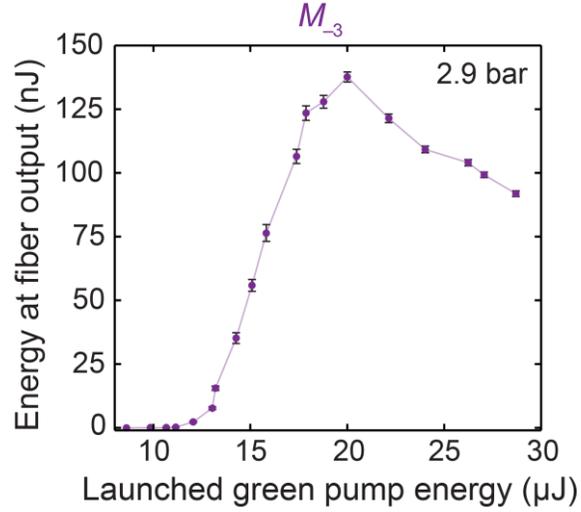

**Fig. S6.**

**Pulse energy of the $M_{-3}$ (199 nm) signal measured at the fiber output for increasing launched green pump energies.** This experimental energy scan was carried out for $M_{00}$~1.43 µJ and when filled with 2.9 bar of hydrogen. $M_{00}$ is defined in main text.

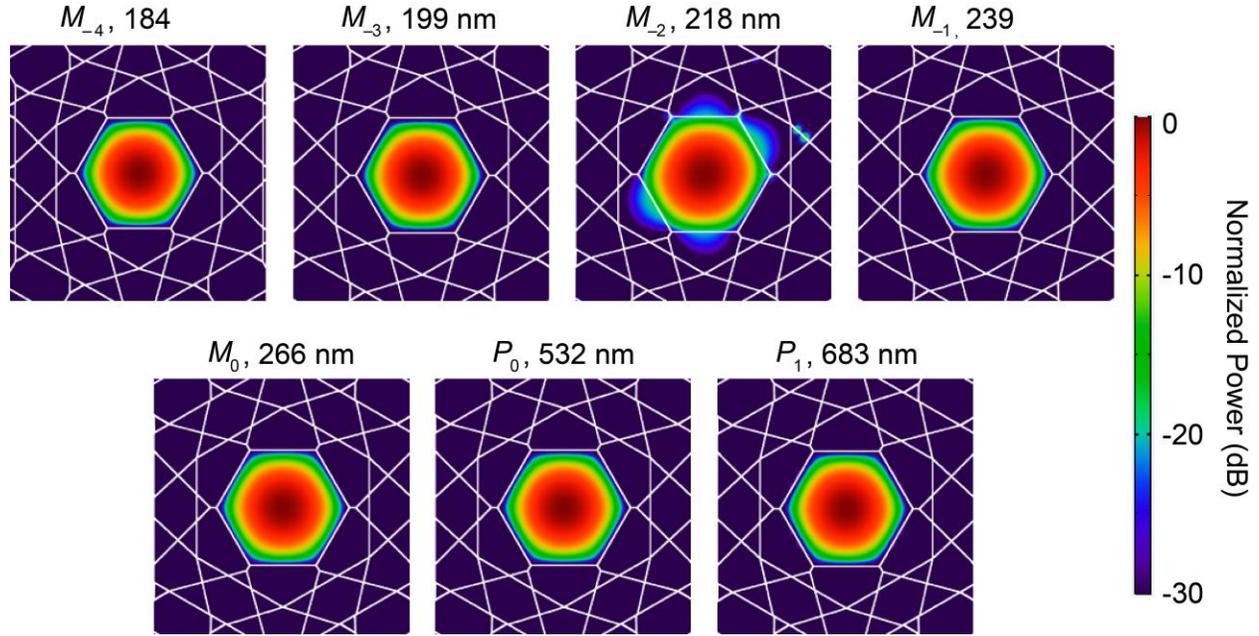

**Fig. S7.**
**Simulated power distribution of the LP$_{01}$-like mode for different wavelengths in a kagomé-style HC-PCF.** The modes are calculated using finite-element modeling of a perfect kagomé-fiber structure with core-wall thickness of ~96 nm, flat-to-flat core-wall diameter of 22 μm and silica glass as the fiber material. Although the $M_{-2}$ signal at 218 nm lies close to the first-order anti-crossing at ~220 nm, only ~1 % of the power is distributed outside the fiber core and hence the real part of its modal refractive index is not strongly affected.

| Pump | | | | | | | Mixing | | | | | | |
|---|---|---|---|---|---|---|---|---|---|---|---|---|---|
| $P_{-3}$ | $P_{-2}$ | $P_{-1}$ | $P_{-0}$ | $P_1$ | $P_2$ | $P_3$ | $M_{-3}$ | $M_{-2}$ | $M_{-1}$ | $M_0$ | $M_1$ | $M_2$ | $M_3$ |
| 4.34 | 2 | 2 | 0.96 | 1.94 | 30 | 20 | 4 | 4 | 4 | 4 | 4 | 4 | 40 |

**Table S1: Numerical loss coefficients in dB/m for the different sidebands travelling in the LP$_{01}$ mode.**

11